\documentclass[%
 reprint,
showpacs,
 amsmath,amssymb,
 aps,
 pra,
 longbibliography,
 floatfix,
 superscriptaddress
]{revtex4-2}
\usepackage{xcolor}
\usepackage{soul}
\usepackage[utf8]{inputenc}	
\usepackage[T1]{fontenc}	
\usepackage[]{graphicx}		
\usepackage{xcolor}
\usepackage{gensymb}
\usepackage{amsmath, mathdots}
\usepackage{bbold}
\usepackage{braket}
\usepackage{comment}
\usepackage[english]{babel}
\usepackage{color}
\usepackage{bm}

\renewcommand{\a}{\hat{a}}
\renewcommand{\b}{\hat{b}}

\newcommand{\q}{\hat{q}}
\newcommand{\p}{\hat{p}}
\renewcommand{\H}{\hat{H}}

\renewcommand{\L}{\hat{L}}

\newcommand{\U}{\hat{U}}

\renewcommand{\Re}{\text{Re}}
\renewcommand{\Im}{\text{Im}}

\begin{document}

\title{Efficient Description of Parametric Amplification of Quantum Pulses}

\author{Victor Rueskov Christiansen}
\email{victorrc@phys.au.dk}
\affiliation{Center for Complex Quantum Systems, Department of Physics and Astronomy, Aarhus University, Ny Munkegade 120, DK-8000 Aarhus C, Denmark}

\author{Klaus Mølmer}
\email[]{klaus.molmer@nbi.ku.dk}
\affiliation{Center for Hybrid Quantum Networks (Hy-Q), Niels Bohr Institute, University of Copenhagen, Blegdamsvej 17, 2100 Copenhagen, Denmark.}

\author{Emanuel Hubenschmid}
\email[]{emanuel.hubenschmid@uni-konstanz.de}
\affiliation{Department of Physics, University of Konstanz, D-78457 Konstanz, Germany}

\begin{abstract}
\noindent 
A single quantum pulse undergoing parametric amplification feeds into at most two pulses in the output. In this work, we present an efficient, analytical method for finding the quantum state of these output modes. Our method applies the amplification to the vacuum rather than to the input state, and subsequently applies a transformed version of the operator that creates the input state from vacuum. Given the input and output pulse mode functions, the method is analytical, and therefore computationally very efficient, and it can be readily generalized to multiple non-vacuum input modes. We exemplify the method by computing the output quantum states resulting from the input of a coherent, a Schrödinger cat, and a single photon input quantum state. We further employ the method to obtain the quantum state in one of the two output modes heralded upon detection of vacuum in the other, least populated, mode.
\end{abstract}
\maketitle

\noindent
\section{Introduction}
Bosonic quantum states are integral to many quantum applications, such as quantum networks \cite{quantum_networks}, quantum computing \cite{photonic_quantum_tech} and sensing \cite{caves, caves2, caves3, sensing}. The description of bosonic radiation traveling in, e.g., wave guides between nodes in open system quantum networks is challenging since the radiation may explore a continuum of frequency modes and require a multi-mode description \cite{temporal_modes, fabre-treps-modes}. In this work, we describe multi-mode quantum states of light generated by a Hamiltonian that is quadratic in the creation and annihilation operators of the bosonic excitations of the system. Such interactions describe, e.g., beam-splitting, mode sorting by spectral filtering, linear attenuation, parametric amplification and squeezing.

Parametric amplifiers are important elements in quantum optics and quantum networks with bosonic carriers of information, generating non-classical and entangled resource states \cite{grid_states, GKP_OPA_singlemode_approx, jeannic_creating_cats, takase_creating_cats, chen_creating_cats_addition, asavanant_creating_cats}, allowing quantum sensing beyond the standard quantum limit \cite{caves, caves2}, and enabling measurement based quantum computing \cite{cluster_state, measurement_based} and  Gaussian boson sampling \cite{gaussian_boson_sampling_yu, gaussian_boson_sampling_hamilton}. Here we analyze the effect of parametric amplifiers on non-vacuum inputs which are of high interest in many applications, for instance in the readout of the quantum states of superconducting qubits \cite{sc_qubits_readout_twpa_macklin, sc_qubits_readout_twpa_qui}, in quantum state tomography \cite{tomography_via_OPA, measurement_phase_sensitive_amplifier,Hubenschmid2022,Hubenschmid2024,Hubenschmid2026}, and in continuous-variable quantum computing \cite{grid_states, GKP_OPA_singlemode_approx, continuous_variables}, where squeezing also offers protection from noise and decoherence \cite{protecting_cat_states_jeannic, protecting_cat_states_park, protecting_cat_states_pan}. In all of these applications, characterization of the quantum state of the output of the parametric amplifier is of critical importance. This is an easy task in the case of single mode operation, but even for a single input pulse mode to a parametric amplifier, the output field will be distributed in an entangled manner over modes that have experienced varying degrees of squeezing. 

It was recently shown that a single quantum pulse undergoing a parametric amplification transformation will at most contribute to the quantum state contents of two, and in some cases only one, pulse modes in the output field, while all other output modes will only have (squeezed) contributions from vacuum input \cite{parametric-amplification-quantum-pulse}. In this work, we offer an analytical approach to find the quantum state in the output modes fed by the input pulse. Our method is efficient for any Fock state superposition or mixed state in the input pulse mode.

The work is structured as follows. In Sec.~\ref{sec:operator-transformation} we introduce the input-output relation of a second-order nonlinear system, and we outline the analytical method for finding the output quantum state in the relevant modes occupied by the input quantum state. In Sec.~\ref{sec:input-decomposition}, we identify the relevant output modes. In Sec.~\ref{sec:squeezed-vacuum-state}, we compute the squeezed state contents of the output modes in case of vacuum input to the amplifier. In Sec.~\ref{sec:output-quantum-state}, we present the formalism to evaluate the output for arbitrary input quantum state, and show specific results for a coherent state, a Schrödinger cat state, and a single photon Fock state. In Sec.~\ref{sec:heralded_quantum_state}, we show that detection of vacuum in one of the two output modes can lead to a purified quantum state in the other mode. In Sec.~\ref{sec:conclusion} we summarize the work and give an outlook on future potential applications of our method.

\begin{figure*}
	\centering
	\includegraphics[width=0.95\textwidth]{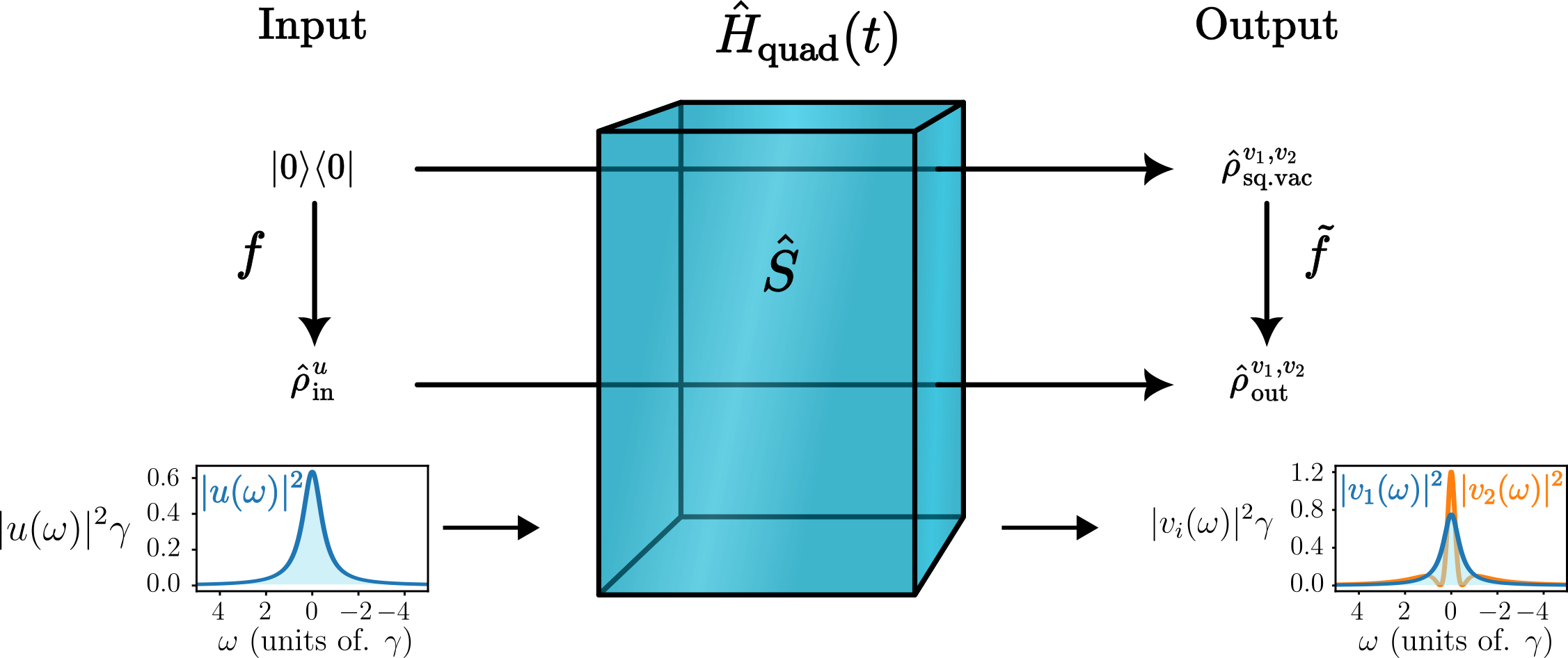}
	\caption{Schematic depiction of the transformation of quantum states and methods discussed in this article. An input state $\hat{\rho}_\textup{in}^u = f(\a_u, \a_u^\dagger) \ket{0}\bra{0} [f(\a_u, \a_u^\dagger)]^\dagger$, generated by applying $f$ to the vacuum, in a single pulsed mode with spectral mode function $u(\omega)$ is incident on a nonlinear system offering frequency conversion and parametric amplification. The nonlinear interaction is described by a Hamiltonian, which may be explicitly time-dependent and is quadratic in its creation/annihilation operators. Only two output modes, $v_1(\omega)$ and $v_2(\omega)$, will be fed by the input quantum state. The lower panels show an input mode and the corresponding state dependent output modes for an optical parametric oscillator with natural decay rate $\gamma$ irradiated by a single photon. The quantum state contents of the output modes may be calculated directly by expanding $u(\omega)$ on the full set of (Bogoliubov) eigenmodes of the transformation or on a judiciously chosen smaller set of modes. In this article we present a more efficient approach that first calculates the transformation of a vacuum input state (upper arrow), and subsequently acts by a transformed version $\tilde{f}$ of the operator $f$ that creates the input state from vacuum.}
	\label{fig:different-methods-diagram}
\end{figure*}

\section{Transformation of field operators by a parametric amplifier} \label{sec:operator-transformation}
We consider pulsed excitations of a bosonic quantum field, propagating in one dimension and being subject to a second-order nonlinearity.
The system may be described by a Hamiltonian that is at most quadratic in its annihilation and creation operators.
Solving the input-output relation, results in a linear relation between the annihilation operator of the output quantum field $\b_\text{out}(\omega)$ and the annihilation and creation operator of the input field $\a_\text{in}(\omega)$, $\a_\text{in}^\dagger(\omega)$.
Since the nonlinear interaction can mediate between different frequencies, the input-output relation
\begin{align} \label{eq:output-input-relation}
    \b_\text{out}(\omega) = \int d\omega' \left[F(\omega, \omega') \a_\text{in}(\omega') + G^*(\omega, \omega') \a_\text{in}^\dagger(\omega') \right],
\end{align}
includes the kernels $F(\omega, \omega')$ and $G(\omega, \omega')$ for transferring between frequencies $\omega$ and $\omega'$. The kernels conserve the bosonic commutation relation of the mode operators. We emphasize that this input-output relation may also describe explicit time-dependent second-order nonlinearities.
An example of such a system is a pulsed optical parametric oscillator (OPO), which is described in Sec.~\ref{sec:output-quantum-state}, and used for the examples in Sec.~\ref{sec:output-quantum-state} and~\ref{sec:heralded_quantum_state}.
The results presented in our work can be generalized to fields propagating in more spatial dimensions~\cite{parametric-amplification-quantum-pulse}.

The inverse transformation can be written as \cite{Braunstein2005}
\begin{align} \label{eq:input-output-relation}
    \a_\text{in}(\omega) = \int d\omega' \left[F^*(\omega', \omega) \b_\text{out}(\omega') - G^*(\omega', \omega) \b_\text{out}^\dagger(\omega') \right].
\end{align}
In the following, we omit the subscripts ``in'' and ``out'' and refer to $\a$-operators as operators acting on the input, and $\b$-operators as operators acting on the output.

We assume the field at the input port of the nonlinear interaction is populated by a single input mode with spectral amplitude $u(\omega)$, the Fourier transform of its temporal wave packet $\tilde{u}(t)$, and we are interested in finding the transformed quantum state at the output port. Our theory can be generalized to multiple modes in the input port, but here we focus on a single mode.

One method to calculate the output quantum state uses a Bogoliubov eigenmode decomposition of $F(\omega, \omega')$ and $G(\omega, \omega')$ to reduce the problem to one of single mode squeezing of these eigenmodes \cite{Wasilewski2006}. This approach is useful in the case where the input occupies only a few of the eigenmodes, while if this is not the case, the expansion on the eigenmodes of a general input state has a complexity of the same order as the boson sampling problem. This complexity occurs when the nonlinearity is used to amplify or up-convert quantum states of radiation across a range of frequencies or temporal modes \cite{Hubenschmid2026}.

Another method achieves an effective description by decomposing the output operator in Eq.~\eqref{eq:output-input-relation} in terms of the input mode and a few additional, orthogonal vacuum input modes. This is the approach taken in Ref. \cite{parametric-amplification-quantum-pulse}, where it is shown that any input quantum state in a single input mode $u$, will at most occupy two modes, $v_1$ and $v_2$, in the output. These modes will also contain parametrically amplified contributions from the vacuum input modes. Therefore, the method provides a resourceful mode decomposition for any given input state to the parametric amplifier.

Both the methods mentioned above rely on ancillary modes for calculating the output quantum state. Here, we utilize the mode decomposition of Ref. \cite{parametric-amplification-quantum-pulse} and provide a method that only includes the physical output modes and no ancillary modes. The resourceful use of modes enables the most efficient calculation of the output quantum state in terms of Hilbert space dimensionality. Therefore, it prevents the computational complexity encountered in the mixing of many modes to become an issue.
As the decomposition in Ref. \cite{parametric-amplification-quantum-pulse} is relevant for the derivations presented in this paper, we summarize the argument leading to the two output modes $v_1$ and $v_2$ in App.~\ref{sec:calculating-v1-v2}. Furthermore, we offer an alternative argument in App.~\ref{app:alternative-two-mode-proof}, and we summarize the decomposition of output modes on a small set of input modes in App.~\ref{sec:output-decomposition}.

Rather than following Ref.~\cite{parametric-amplification-quantum-pulse} and developing the theory around the output operator in Eq. \eqref{eq:output-input-relation}, in this work we take our starting point in a decomposition of the input operator in Eq. \eqref{eq:input-output-relation}. Since we know that the non-vacuum input mode leads to output in only two relevant output modes, we are able to expand the input annihilation operator $\a_u$ as
\begin{align} \label{eq:general-input-output-relation}
    \a_u = A \b_{v_1} + B \b_{v_1}^\dagger + C \b_{v_2} + D \b_{v_2}^\dagger,
\end{align}
with coefficients $A, B, C$ and $D$ (see Sec. \ref{sec:input-decomposition}).

If the input state can be expressed as $\ket{\psi_u} = f(\a_u, \a_u^\dagger) \ket{0}$ for some function $f$, we can use \eqref{eq:general-input-output-relation} to find the corresponding output quantum state. Writing the transformation of the nonlinear parametric amplifier as a unitary transformation $\U$, we have 
\begin{align}
    \U \ket{\psi_u} = \U f(\a_u, \a_u^\dagger) \U^\dagger \U \ket{0}.
\end{align}
The evaluation of  $\U f(\a_u, \a_u^\dagger) \U^\dagger = \tilde{f}(\b_{v_1}, \b_{v_1}^\dagger, \b_{v_2}, \b_{v_2}^\dagger)$ merely entails replacing arguments $\a_u$ and $\a_u^\dagger$ in $f$ by their expressions following from Eq. \eqref{eq:general-input-output-relation}. The resulting operator should then be applied on the squeezed multimode vacuum state $\ket{\xi_i} = \U \ket{0}$. In summary, the input state transforms into the output state as given by 
\begin{align}
    f(\a_u, \a_u^\dagger) \ket{0} \rightarrow \tilde{f}(\b_{v_1}, \b_{v_1}^\dagger, \b_{v_2}, \b_{v_2}^\dagger) \ket{\xi_i}.
\end{align}
While the input states we consider here are pure, the formalism also applies to mixed input states, by replacing $\ket{0}$ with $\ket{0}\bra{0}$ and $f(\a_u, \a_u^\dagger)$ by a super operator $\mathcal{F}(\a_u, \a_u^\dagger)$, which creates the desired mixed state under the action upon $\ket{0}\bra{0}$.

The two different paths to obtain the output quantum state are illustrated in Fig.~\ref{fig:different-methods-diagram}, where the method applied in this article distinguishes itself by first evaluating the effect of the amplifier on the vacuum state and thereafter performing an additional operation to yield the desired outcome for a non-trivial initial state. Details of the scheme will be presented in the following sections. We remind that while we focus here on a single input pulse mode and the quantum state in two output modes $v_1$ and $v_2$, the method is more general and can be applied also with more input and output modes.

\section{Decomposition of the input mode} \label{sec:input-decomposition}
It was shown in \cite{parametric-amplification-quantum-pulse} that a quantum state occupying a pulse of radiation $u(\omega)$ propagating along a quantum channel incident on a parametric amplifier will in general only contribute to the population of two modes, $v_1(\omega)$ and $v_2(\omega)$, of the output quantum field. It was also shown that if the input quantum state obeys the relation 
\begin{align} \label{eq:single-mode-condition}
    \braket{\a_u^\dagger \a_u} = |\braket{\a_u \a_u}|,
\end{align}
only one pulse in the output, $v(\omega)$, will be populated by the input state contents. In the following we consider the decomposition of the input mode, first for the case where only a single output mode is occupied by the input, and then for the general case of two populated output modes.

\subsection{The single-mode transformation}
We consider first the case where the input quantum state obeys the single-mode condition in Eq.~\eqref{eq:single-mode-condition}. Examples of states for which this holds are the coherent states $\ket{\alpha}$ and Schrödinger cat states of the form $\ket{\alpha} \pm i\ket{-\alpha}$. In Ref.~\cite{parametric-amplification-quantum-pulse} it is found that the output mode occupied by the input quantum state is given by
\begin{equation} \label{eq:v-single-mode}
    v(\omega) = \frac{\alpha \zeta_u f_u(\omega) + \alpha^* \xi_u g_u(\omega)}{k},
\end{equation}
where $\alpha = \sqrt{\braket{\a_u \a_u}}$ and  
\begin{align}
    f_u(\omega) &= \frac{1}{\zeta_u} \int d\omega' F(\omega, \omega') u(\omega') \label{eq:f_u}, \\
    g_u(\omega) &= \frac{1}{\xi_u} \int d\omega' G^*(\omega, \omega') u^*(\omega')\label{eq:g_u}.
\end{align}
The constants $\zeta_u$, $\xi_u$ and $k$ ensure the normalization of $f_u(\omega)$, $g_u(\omega)$ and $v(\omega)$, respectively, where $k$ is given by
\begin{align}
\begin{split}
    k^2 &= |\alpha|^2(\zeta_u^2 + \xi_u^2) + 2\zeta_u \xi_u \Re(\alpha^2 \braket{g_u, f_u}).
\end{split}
\end{align}
We use the inner product notation, $\braket{f, g} = \int d\omega f^*(\omega)g(\omega)$ for mode overlap integrals and define the orthogonality of mode functions accordingly. The annihilation operator for the input mode $u$ can be expanded as
\begin{align}
\begin{split}
    \a_u 
    &= \int d\omega d\omega' u^*(\omega) F^*(\omega', \omega) \b_\textup{out}(\omega') \\&-  \int d\omega d\omega' u^*(\omega) G^*(\omega', \omega) \b_\textup{out}^\dagger(\omega') \\
    &= \zeta_u \b_{f_u} - \xi_u \b_{g_u}^\dagger.
\end{split}
\end{align}
We note that $f_u$ and $g_u$ are not orthogonal but we can expand them on the output mode $v(\omega)$ and an orthogonal component that can be found by a Gram-Schmidt procedure,
\begin{align}
\begin{split}
    h(\omega) &= \frac{f_u(\omega) - \braket{v, f_u} v(\omega)}{\sqrt{1 - |\braket{v, f_u}|^2}}.
\end{split}
\end{align}
This yields
\begin{align}
\begin{split}
    \a_u &= \zeta_u \braket{f_u, v} \b_v + \zeta_u \braket{f_u, h} \b_h^\dagger \\
    &- \xi_u \braket{v, g_u} \b_v^\dagger - \xi_u \braket{h, g_u} \b_h^\dagger.
\end{split}
\end{align}
From Eq. \eqref{eq:v-single-mode}, it can be shown that $\xi_u\braket{v, g_u} = \zeta_u \braket{f_u, v} - \alpha/k$, and it follows that $\xi_u \braket{h, g_u} = - \zeta_u \alpha/\alpha^* \braket{h, f_u}$. Finally, it can be seen that $\braket{f_u, h} = \sqrt{1 - |\braket{f_u, v}|^2}$, and we arrive at writing the input mode in terms of the output modes as
\begin{align} \label{eq:single-mode-operator}
\begin{split}
    \a_u &= \zeta_u \braket{f_u, v} \b_v - \left(\zeta_u\braket{f_u, v} - \frac{\alpha}{k}\right) \b_v^\dagger \\
    &+ \zeta_u \sqrt{1-|\braket{f_u,v}|^2} \left(\b_h + \frac{\alpha}{\alpha^*} \b_h^\dagger\right).
\end{split}
\end{align}
At first glance, Eq.~\eqref{eq:single-mode-operator} seems to contradict that the input-dependent part of the output quantum state resides in a single pulsed mode. Taking a closer look at the quantum states, for instance the coherent states, we can explicitly transform the corresponding displacement operator,
\begin{align} \label{eq:displacement-operator}
\begin{split}
    D_u(\alpha) &= \exp\left(\alpha \a_u^\dagger - \alpha^* \a_u\right) \\
    &= \exp\left(\left(2\zeta_u \textup{Re}(\alpha\braket{v, f_u}) - \frac{|\alpha|^2}{k}\right) (\b_v^\dagger - \b_v) \right),
\end{split}
\end{align}
and we see that the $\b_h$ components disappear from the transformation. The $\a_u$-operator only acts on one mode in the output field as expected.

While it is common in quantum optics to choose the mode basis in order to diagonalize the Hamiltonian, we note that the minimal mode decomposition in Eq.~\eqref{eq:single-mode-operator} depends on the state populating the input mode, as indicated by the explicit dependence of $\alpha$ in the mode expansion coefficients $\zeta_u$, $\xi_u$.

\subsection{The two-mode transformation}
We now extend the analysis to the case of general quantum states occupying a single mode input pulse $u$. In this case, the relevant quantum state populates two modes at the output. These output modes can be found by considering the autocorrelation function \cite{parametric-amplification-quantum-pulse}
\begin{align} \label{eq:g1}
\begin{split}
    &g_{1}(\omega,\omega') = \braket{\b^\dagger(\omega) \b(\omega')} \\
    &= \braket{\a_{u}^{\dagger}\a_{u}} \zeta_u^2 f_u^*(\omega) f_u(\omega') + \braket{ \a_{u}^{\dagger}\a_{u}^{\dagger}} \zeta_u \xi_u f_u^*(\omega) g_u(\omega') \\ &+\braket{\a_{u}\a_{u}} \zeta_u\xi_u g_u^*(\omega) f_u(\omega') + \braket{\a_{u}^{\dagger}\a_{u}} \xi_u^2 g_u^*(\omega) g_u(\omega') \\ &+\int d\omega''G(\omega,\omega'')G^{*}(\omega',\omega'').
\end{split}
\end{align}
The first four terms in the expression for the $g_1$-function depend on the input quantum state, while the last term arises from vacuum input. Since the terms that depend on the input quantum state are given by two functions $f_u$ and $g_u$, as defined in eq.~\eqref{eq:f_u} and \eqref{eq:g_u}, the autocorrelation function can be decomposed into at most two orthogonal modes, $v_1$ and $v_2$ \cite{parametric-amplification-quantum-pulse} (see Appendix \ref{sec:calculating-v1-v2}). Decomposing the input-output relation in Eq. \eqref{eq:input-output-relation} reveals
\begin{align}
    \a_u &= \int d\omega' f_u^*(\omega') \b_\textup{out}(\omega') -  \int d\omega' g_u(\omega') \b_\textup{out}^\dagger(\omega'),
\end{align}
and since $f_u$ and $g_u$ can be expressed using the two orthogonal eigenmodes $v_1$ and $v_2$ of the autocorrelation function $g_1(\omega_1, \omega_2)$, we can rewrite the input-mode operator as
\begin{align} \label{eq:two-mode-operator}
\begin{split}
    \a_u &= \zeta_u \braket{f_u, v_1} \b_{v_1} - \xi_u \braket{v_1, g_u} \b_{v_1}^\dagger \\
    &+ \zeta_u \braket{f_u, v_2} \b_{v_2} - \xi_u \braket{v_2, g_u} \b_{v_2}^\dagger.
\end{split}
\end{align}
In this case, it is evident that the input quantum state in  mode $\a_u$ only contributes to the output quantum state in the two output modes $\b_{v_1}$ and $\b_{v_2}$. Explicit expressions for the modes $v_1(\omega)$ and $v_2(\omega)$ are given in appendix \ref{sec:calculating-v1-v2}. As for the single-mode transformation in the previous section, the two modes, $v_1(\omega)$ and $v_2(\omega)$, depend on the input quantum state of the mode $u(\omega)$, since they diagonalize the (state-dependent) $g_1$-correlation function.

\section{The output squeezed vacuum state} \label{sec:squeezed-vacuum-state}
\begin{figure*}
    \centering
    \includegraphics[width=0.9\textwidth]{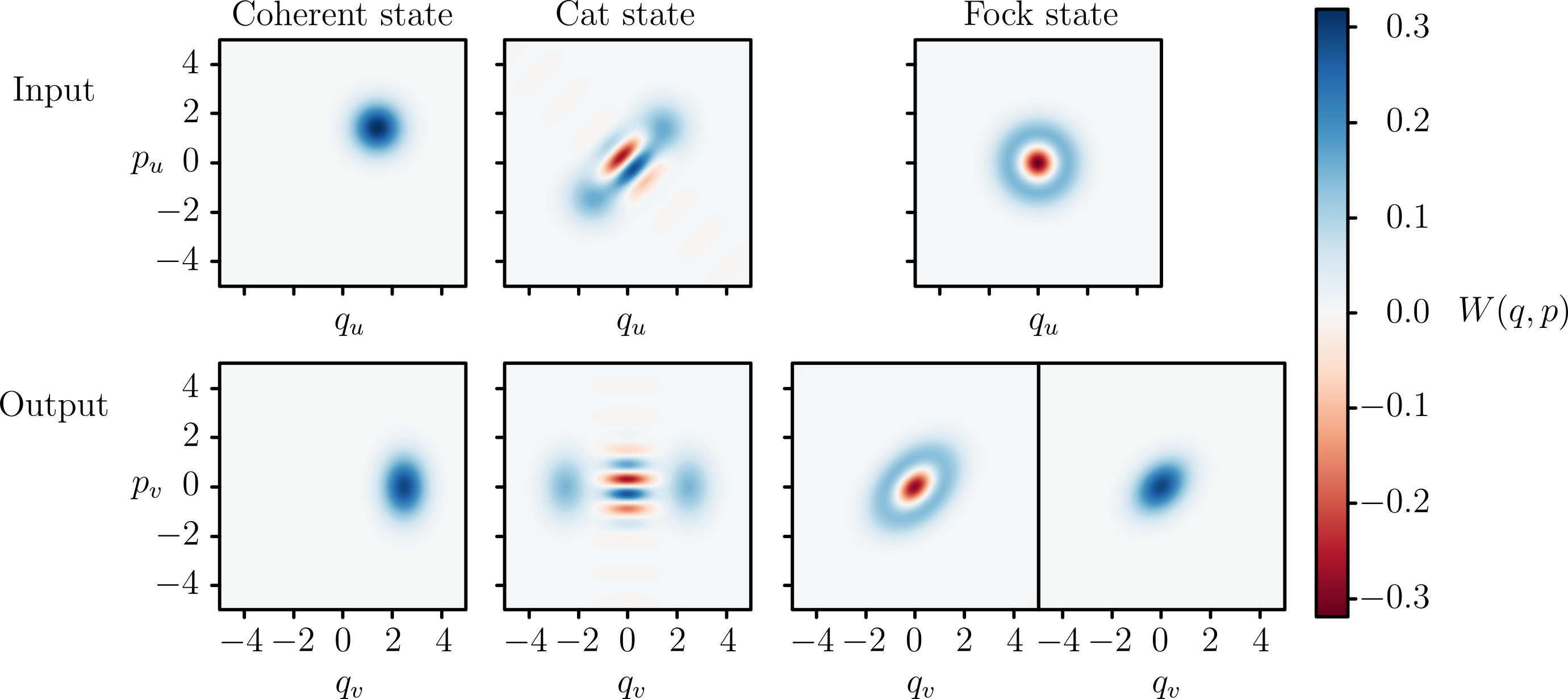}
    \caption{Wigner function of the three example input states, coherent $\ket{\alpha}$, cat $(\ket{\alpha} + \textup{i} \ket{-\alpha})/\sqrt{2}$, and Fock state $\ket{n}$, and their corresponding output states after transformation by an optical parametric oscillator (OPO). The OPO parameters and the input mode function $u(\omega)$ are the same for all examples and are specified in Sec.~\ref{sec:output-quantum-state} and Sec.~\ref{sec:heralded_quantum_state} with the OPO detuning $\Delta = 0$ and pump strength $\xi = 0.1\textup{i} \gamma$ and the frequency width of the pulse $\Gamma = \gamma$, where $\gamma$ is the natural decay rate of the OPO cavity. The input mode, $u(\omega)$, is shown in Fig. \ref{fig:different-methods-diagram}, along with the two corresponding output modes, $v_1(\omega), v_2(\omega)$, for a single photon input state. Both the coherent and cat input state have a parameter $\alpha = 1 + \textup{i}$, while the input Fock state is a single photon state $n=1$. Since the Fock state has two output modes, their Wigner functions are plotted side by side, with $v_1(\omega)$ to the left, and $v_2(\omega)$ to the right. The phase space arguments $q$ and $p$ correspond to the quadrature observables, $\q = (\a + \a^\dagger)/\sqrt{2}$ and $\p = (\a - \a^\dagger)/\textup{i}\sqrt{2}$, defined in the main text.}
    \label{fig:example_squeezed_states}
\end{figure*}

Computing the output quantum state for the case of vacuum in all input modes, including mode $u$, serves as starting point to calculate the transformation of more intricate quantum states. The method is general and will be presented here for any number of input modes and a corresponding (potentially different) number of relevant output modes.
We first note that the parametrically amplified vacuum is a Gaussian state and it is described completely by the first and second moments of the quadrature operators, $\q_i = (\a_i + \a_i^\dagger)/\sqrt{2}$ and $\p_i = (\a_i - \a_i^\dagger)/\textup{i}\sqrt{2}$. These form the displacement vector $\mathbf{\bar{r}} = \braket{\mathbf{\hat{r}}}$, where $\mathbf{\hat{r}} = (\hat{q}_1, \dots, \hat{q}_n, \hat{p}_1, \dots, \hat{p}_n)^T$, and the covariance matrix $\bm{\sigma}$ with entries $\sigma_{kl} = \braket{\{\hat{r}_k, \hat{r}_l\}} - 2\braket{\hat{r}_k}\braket{ \hat{r}_l}$ \cite{Weedbrook2012,Adesso2014}. 

\subsection{The transformed vacuum in phase space} \label{sec:gaussian-variables-output-state}
We start by calculating the displacement vector and the covariance matrix of the general $N$-mode squeezed vacuum output state of the parametric amplifier. We express the transformation in terms of general matrices (see App.~\ref{sec:output-decomposition} for the entries of $E$ and $F$ in the single-mode and two-mode cases),
\begin{align} \label{eq:input-output-matrix}
    \mathbf{\b} = \begin{pmatrix}
        E & F \\ F^* & E^*
    \end{pmatrix} \mathbf{\a}
\end{align}
where we define vectors consisting of input and output operators $\mathbf{\b} = (\b_1, \dots, \b_N, \b_1^\dagger, \dots, \b_N^\dagger)^T$ and $\mathbf{\a} = (\a_1, \dots, \a_M, \a_1^\dagger, \dots, \a_M^\dagger)^T$ with $N$ output modes and $M$ input modes, and $E$ and $F$ are $N\times M$ matrices describing the input-output relation of these operators. We use $E^*$ to denote complex conjugation of each matrix element of $E$, and $E^\dagger$ to denote complex conjugation and transposition, $E^\dagger = (E^*)^T$. The full multi-mode transformation is invertible, {\it cf}., Eq.~\eqref{eq:output-input-relation} and Eq.~\eqref{eq:input-output-relation}, but we may be interested in a truncation of the problem to certain modes of interest, and the input-output relation in Eq.~\eqref{eq:input-output-matrix} between input operators $\mathbf{\a}$ and output operators $\mathbf{\b}$ does not need to be invertible, and may, e.g., entail, $N\neq M$. 

In the quadrature basis, $\mathbf{\q_\textup{in}}$ and $\mathbf{\p_\textup{in}}$ are vectors with the $\q_i = (\a_i + \a_i^\dagger)/\sqrt{2}$ and $\p_i = (\a_i - \a_i^\dagger)/\textup{i}\sqrt{2}$ quadratures as entries and similar for $\mathbf{\q_\textup{out}}$ and $\mathbf{\p_\textup{out}}$ in terms of $\b_i$. In this basis, we find the covariance matrix of the output as (see App.~\ref{sec:vacuum-state-covariance} for a derivation),
\begin{align}
    \bm{\sigma} = I + 2 \begin{pmatrix}
        \text{Re}(EF^T + FF^\dagger) & \text{Im}(EF^T - FF^\dagger) \\
        \text{Im}(EF^T + FF^\dagger) & \text{Re}(FF^\dagger - EF^T)
    \end{pmatrix},
\end{align}
while the displacement vector is the zero-vector. 

\subsection{The density matrix of the squeezed vacuum state}
The expansion in the Fock basis of the density matrix of a Gaussian state with covariance matrix $\bm{\sigma}$ and displacement vector $\mathbf{\bar{r}}$ is given by multidimensional Hermite polynomials $H^{\{\mathbf{R}\}}_{\mathbf{m}, \mathbf{n}}$ \cite{gaussian-density-matrix},
\begin{align} \label{eq:gaussian-density-matrix}
    \hat{\rho} = P_0 \sum_{\mathbf{m},\mathbf{n}} \frac{H^{\{\mathbf{R}\}}_{\mathbf{m}, \mathbf{n}}(\mathbf{y})}{\sqrt{\mathbf{m}! \mathbf{n}!}} \ket{\mathbf{m}}\bra{\mathbf{n}},
\end{align}
where $\mathbf{m} = (m_1, \dots m_N)$, $\mathbf{n} = (n_1, \dots, n_N)$, and $\mathbf{m}!\mathbf{n}! = m_1!\dots m_N! n_1! \dots n_N!$, while we have defined the matrix $\mathbf{R}$ and vector $\mathbf{y}$ as 
\begin{align}
    \mathbf{R} &= \mathbf{U}^\dagger (\mathbf{I}_{2N} - \bm{\sigma})(\mathbf{I}_{2N} + \bm{\sigma})^{-1} \mathbf{U}^*, \\
    \mathbf{y} &= 2\mathbf{U}^T (\mathbf{I}_{2N} - \bm{\sigma})^{-1} \mathbf{\bar{r}},
\end{align}
where $\mathbf{U}$ is
\begin{align}
    \mathbf{U} = \frac{1}{\sqrt{2}} \begin{pmatrix}
        \mathbf{I}_N & \mathbf{I}_N \\
        -i \mathbf{I}_N & i \mathbf{I}_N
    \end{pmatrix},
\end{align}
and $\mathbf{I}_N$ denotes the $N \times N$ identity matrix. The factor $P_0$ denotes the probability to have zero photons in all modes and is given by
\begin{align}
    P_0 = \frac{\exp\left(-\mathbf{\bar{r}}^T (\bm{\sigma}+\mathbf{I}_{2N})^{-1} \mathbf{\bar{r}}\right)}{\sqrt{\text{det}[(\bm{\sigma} + \mathbf{I}_{2N}) / 2]}}.
\end{align}


The multidimensional Hermite polynomials can be calculated using recursion, \cite{multidimensional-hermite-polynomials, hermite-pol-in-q-optics, the-walrus}. We have used the implementation of \texttt{the-walrus} package for \texttt{Python} \cite{the-walrus} as part of this work, see \cite{code}.

Since the general output state is constructed by applying the function $f$ to the vacuum state, for now we only need to consider transformations with the vacuum as input, in which case the output state $\hat{\rho}_\textup{vac, out}$ has a vanishing displacement vector, $\mathbf{\bar{r}} = \mathbf{0}$. This implies that the multidimensional Hermite polynomial shall be evaluated at $\mathbf{y} = \mathbf{0}$,
\begin{align} \label{eq:vacuum-output-state}
    \hat{\rho}_\textup{vac, out} = P_0 \sum_{\mathbf{m},\mathbf{n}} \frac{H^{\{\mathbf{R}\}}_{\mathbf{m}, \mathbf{n}}(\mathbf{0})}{\sqrt{\mathbf{m}! \mathbf{n}!}}\ket{\mathbf{m}}\bra{\mathbf{n}}.
\end{align}
Having found, the density matrix of the squeezed thermal output state resulting from the transformation of the vacuum input state, we now turn to the calculation of the output state for any input quantum state.

\section{The output quantum state} \label{sec:output-quantum-state}
To compute the output state for any non-trivial input quantum state we assume the construction of such a state by the action of a sequence of creation and annihilation operators, parametrized by a function of these operators $f(\a_u, \a_u^\dagger)$, on the vacuum state. This yields an input state written here as a density matrix
\begin{align}\label{eq:input_quantum_state} 
    \hat{\rho}_\textup{in} = f(\a_u, \a_u^\dagger) \ket{0} \bra{0} \left[f(\a_u, \a_u^\dagger)\right]^\dagger.
\end{align}
By replacing $f$ with an appropriate super operator, extension of the theory to mixed input states is straightforward. We must, anyway, apply a density matrix description, because the squeezed vacuum state $\hat{\rho}_\textup{vac,out}$ is entangled across all relevant and irrelevant modes, and its restriction to the one or two modes of interest results in a mixed Gaussian state \cite{Onoe2022,Hubenschmid2026}.
We can now obtain the output state for any given input quantum state by replacing $f(\a_u, \a_u^\dagger)$ and $\hat{\rho}_\textup{vac,in}=\ket{0}\bra{0}$ with the expressions for $\tilde{f}(\b_{v_1}, \b_{v_1}^\dagger, \b_{v_2}, \b_{v_2}^\dagger)$ and $\hat{\rho}_\textup{vac,out}$,
\begin{align}
\begin{split}
    \hat{\rho}_\textup{out} &= \tilde{f}(\b_{v_1}, \b_{v_1}^\dagger, \b_{v_2}, \b_{v_2}^\dagger) \hat{\rho}_\textup{vac,out} \left[\tilde{f}(\b_{v_1}, \b_{v_1}^\dagger, \b_{v_2}, \b_{v_2}^\dagger)\right]^\dagger.\label{eq:output_quantum_state}
\end{split}
\end{align}
For the convenience of the reader, we have implemented all steps applied in our method in a Python library \cite{code} using the \texttt{QuTiP} package for \texttt{Python} \cite{qutip1, qutip2}. In the following, we demonstrate application of the above equation using the examples of a coherent state, a Schrödinger cat state and a Fock state in a single input mode $u(\omega)$. The code needed to generate the examples can be found in the accompanying repository \cite{code}.

For the analysis, we assume an optical parametric oscillator (OPO) Hamiltonian,
\begin{align} \label{eq:opo}
    \H_\textup{OPO} = \Delta \a^\dagger \a + \frac{i}{2}\left[\xi(t) (\a^\dagger)^2 - \xi^*(t) \a^2\right],
\end{align}
where $\a$ is the operator associated with annihilating one quantum of radiation in the cavity mode, $\Delta$ is the detuning between the oscillator frequency and the (classical) optical pump frequency, and $\xi$ represents the pump power, which may be constant or pulsed. The Hamiltonian describes a cavity mode, which is, in addition, coupled to the input and output field with the dissipation operator
\begin{align}
    \L = \sqrt{\gamma} \a,
\end{align}
where $\gamma$ is the cavity photon loss rate, which couples the single mode of the cavity to the continuum of modes outside the cavity. 

For a constant pump $\xi(t) = \xi$, this yields an example of the general input-output relation in Eq.~\eqref{eq:output-input-relation} where $F$ and $G$ are given by \cite{quantum_noise}
\begin{align}
    F(\omega, \omega') &= - \frac{(\frac{\gamma}{2} - i\Delta)^2 + \omega^2 + |\xi|^2}{\Delta^2 + (\frac{\gamma}{2} - i \omega)^2 - |\xi|^2} \delta(\omega-\omega'), \\
    G(\omega, \omega') &= - \frac{\xi^* \gamma}{\Delta^2 + (\frac{\gamma}{2} - i \omega)^2 - |\xi|^2} \delta(\omega + \omega'),
\end{align}
where $\delta(\omega)$ is the Dirac delta function. Here we have assumed a constant pump strength for simplicity, but the general input-output relation for a time-dependent pump can be found in Ref.~\cite{parametric-amplification-quantum-pulse}.

\subsection{Squeezed coherent state}
If the input is in a coherent state $\ket{\alpha}_u$, we know from the condition in Eq. \eqref{eq:single-mode-condition} that the output only occupies a single mode. We can find the output quantum state in that single output mode by applying Eq. \eqref{eq:displacement-operator} to the squeezed vacuum state in Eq. \eqref{eq:vacuum-output-state}, following the above procedure with $f(\hat{a}_u, \hat{a}_u^\dagger)=D_u(\alpha)$. Alternatively, since the input coherent state is a Gaussian state, we can also find the output squeezed coherent state directly by applying the same method as we did for the squeezed vacuum state and plot the Wigner function directly, see the left hand panels of Fig.~\ref{fig:example_squeezed_states}. The squeezed coherent state has a covariance matrix equal to the one of transformed vacuum, but a non-zero displacement vector. Both are given in App.~\ref{sec:vacuum-state-covariance}. The generating code can be found in \cite{code}.

\subsection{Squeezed Schrödinger cat state}
We consider a cat state as the input quantum state of the form
\begin{align}
    \ket{\textup{cat}}_u = \frac{\ket{\alpha}_u + i \ket{-\alpha}_u}{\sqrt{2}},
\end{align}
which obeys Eq. \eqref{eq:single-mode-condition} and therefore only occupies a single output mode. We find the output state in the mode $v$, generated from the cat state at the input, using the method described in the previous sections. First we note that
\begin{align}
    \ket{\textup{cat}}_u = f(\a_u, \a_u^\dagger) \ket{0} = \frac{D_u(\alpha) +i D_u(-\alpha)}{\sqrt{2}} \ket{0},
\end{align}
which means that we can find the output density matrix in the $v$-mode of the transformed cat state by using Eq. \eqref{eq:output_quantum_state} where we insert the vacuum density matrix from Eq. \eqref{eq:vacuum-output-state} and the displacement operator acting on the output state from Eq. \eqref{eq:displacement-operator}. We find the operator $\tilde{f}(\b_{v}, \b_{v}^\dagger)$ is given by
\begin{align}
    \tilde{f}(\b_{v}, \b_{v}^\dagger) = \frac{D_v(\alpha_v) + iD_v(-\alpha_v)}{\sqrt{2}},
\end{align}
with
\begin{align}
    \alpha_v = 2\zeta_u \Re(\alpha \braket{v, f_u}) - \frac{|\alpha|^2}{k}.
\end{align}
The resulting Fock state representation is found numerically and we plot its Wigner function in the middle column of Fig. \ref{fig:example_squeezed_states}. The generating procedure can be found in \cite{code}.

\subsection{Squeezed Fock state}
Next we consider an input Fock state which will in general occupy two modes in the output. We calculate the output quantum state using the same method as for the cat state. An $n$-photon Fock state in the $u$-mode is given by
\begin{align}
    \ket{n}_u = \frac{(\a_u^\dagger)^n}{\sqrt{n!}}\ket{0},
\end{align}
where we immediately identify $f(\a_u, \a_u^\dagger)$. We insert $\a_u$ in terms of output operators for the two mode case, Eq. \eqref{eq:two-mode-operator}, into $f(\a_u, \a_u^\dagger)$ to find $\tilde{f}$,
\begin{align}
\begin{split}
    \tilde{f} = \frac{1}{\sqrt{n!}} &\left(\zeta_u \braket{v_1, f_u} \b_{v_1}^\dagger - \xi_u \braket{g_u, v_1} \b_{v_1}\right. \\
    &\left.+ \zeta_u \braket{v_2, f_u} \b_{v_2}^\dagger - \xi_u \braket{g_u, v_2} \b_{v_2}\right)^n.
\end{split}
\end{align}
We transform the squeezed vacuum state from Eq.~\eqref{eq:vacuum-output-state} as in Eq.~\eqref{eq:output_quantum_state}. With this, we find the squeezed Fock state in the joint output modes $v_1$ and $v_2$. We obtain their Fock state representations numerically and plot their Wigner functions side by side in the right hand panels in Fig. \ref{fig:example_squeezed_states}, where the left panel is the Wigner function of the most occupied $v_1(\omega)$-mode. The generating code can be found in \cite{code}.

\section{Purification of the output state using heralding}\label{sec:heralded_quantum_state}
\begin{figure*}
    \centering
    \includegraphics[width=0.9\linewidth]{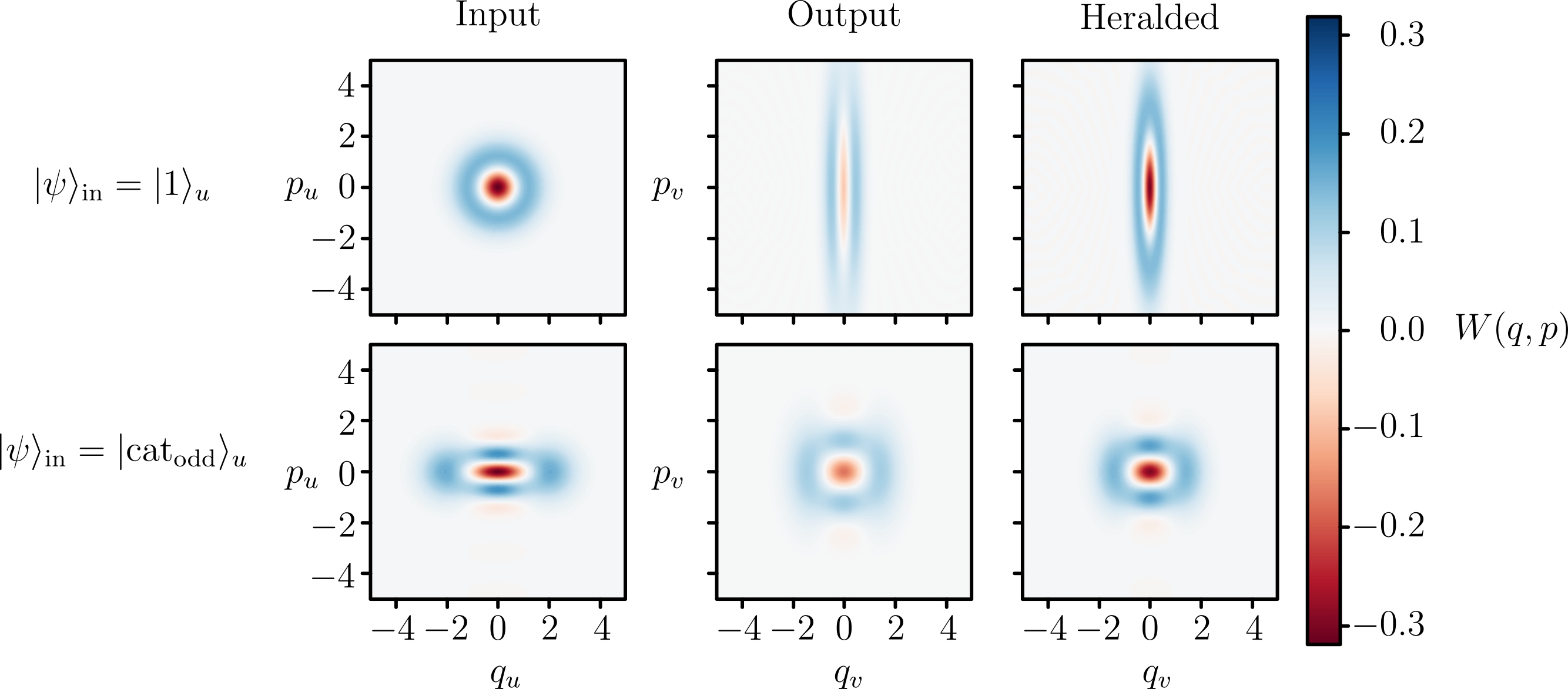}
    \caption{The output quantum state generated from the input of either a single photon state (upper row) or an odd Schrödinger cat state (lower row) residing in the primary mode, $v_1(\omega$). The Wigner functions of the input quantum state are shown in the left column, the output quantum state in the primary $v_1$-mode after tracing out the secondary mode in the middle column, and the quantum state of the primary mode after detection of vacuum in the secondary $v_2$-mode in the right column. The Wigner negativity and purity of the primary mode is enhanced after the detection. For the single photon input state, the purity of the output state is increased from $\gamma_\textup{fock} \approx 0.75$ to $\gamma_\textup{fock} \approx 0.97$ under detection of vacuum, which occurs with a probability $p_{0, \textup{fock}} \approx 52 \%$. For the cat state input the output state has its purity increased from $\gamma_\textup{cat} \approx 0.75$ to $\gamma_\textup{cat} \approx  0.93$ under detection of vacuum with probability $p_{0, \textup{cat}} \approx 43 \%$. We have here considered an open optical parametric oscillator (OPO) with natural decay rate $\gamma$ driven at strength $\xi = -0.75\gamma$ as the quantum system that generates the output quantum states (see Sec.~\ref{sec:output-quantum-state}). The input mode function is a one-sided exponential with width $\Gamma = 2\gamma$ (see Sec.~\ref{sec:heralded_quantum_state}). The phase space arguments $q$ and $p$ correspond to the quadrature observables, $\q = (\a + \a^\dagger)/\sqrt{2}$ and $\p = (\a - \a^\dagger)/\textup{i}\sqrt{2}$, defined in the main text.}
    \label{fig:heralded_quantum_state}
\end{figure*}
To demonstrate the merits of the formalism presented above, we will consider the purification of the output squeezed quantum state, originating from a single-mode input state in mode $u(\omega)$, by detecting vacuum in one of the output modes. From the analysis in the previous sections, we know that the output state will in general occupy two output modes, which we describe by the eigenmodes of the autocorrelation function, $v_1(\omega)$ and $v_2(\omega)$. We consider as input a single photon state and an optical odd Schrödinger cat state, $\ket{\textup{cat}_\textup{odd}} \propto \ket{\alpha} - \ket{-\alpha}$, which will both lead to population in two output modes.
Methods to prepare and manipulate cat states of light are discussed in \cite{takase_creating_cats, jeannic_creating_cats, asavanant_creating_cats, nakamura-cat, ourjoumtsev-cat, polzik-cat}.

We show that detection of vacuum in the least populated output mode, $v_2(\omega)$, will improve the purity and Wigner negativity of quantum state in the most populated output mode, $v_1(\omega)$. This detection can therefore be used to generate more pure and useful squeezed quantum states using an open system parametric amplifier.

We assume the input single photon Fock state is generated with a quantum dot system. Quantum dot systems can be excited and deterministically emit high fidelity single photon states \cite{quantum_dots, single_photon_source} with a mode shape approximately given by an exponential $\tilde{u}(t) = \sqrt{\Gamma}\exp{(-\Gamma t /2)}$ for $t \geq 0$ and $\tilde{u}(t) = 0$, where $\Gamma$ is the decay rate of the emitter. In the frequency domain, this mode shape is
\begin{align} \label{eq:u}
    u(\omega) = \sqrt{\frac{\Gamma}{2\pi}} \frac{i\omega + \Gamma / 2}{\omega^2 + \Gamma^2 / 4}.
\end{align}

For the Schrödinger cat state as input, we assume a similar mode function, which may originate from creating the cat state in an optical cavity, and then emit it by turning on a linear output coupling after the cat state has been generated. This procedure may be optimized through emission while creating the cat~\cite{nakamura-cat}, but here we assume the former procedure is used, for sake of simplicity.

We now calculate the output quantum state of an OPO seeded by the input quantum state in this mode using the method presented in this work. For the example, we choose the decay rate of the single photon source to be $\Gamma = 2\gamma$ and $\xi = -0.75\gamma$ for the OPO in Eq.~\eqref{eq:opo}, with no detuning, $\Delta = 0$. We assume a detection of the number of photons in the second mode $v_2(\omega)$ yields a zero-photon result, and apply the renormalizing projective operator onto this subspace given by $\hat{P} = \hat{I}_{v_1} \otimes \ket{0}\bra{0}_{v_2} / \sqrt{p_0}$, where $p_0$ is the probability of detecting 0 photons. The detection of a single mode in the output may be enabled through a sum-frequency generation protocol~\cite{silberhorn-sum-frequency-generation}.

The Wigner functions of the input and output quantum states for the input in a single photon state and an odd Schrödinger cat state with amplitude $|\alpha|^2 = 2$ are shown in Fig.~\ref{fig:heralded_quantum_state}. We show the output quantum state both before and after heralding on vacuum in the second output mode. For both input states, the purity of the output state after the OPO is $\gamma_\textup{out} \approx 0.75$. For an input single photon state, the purity after detecting vacuum in mode $v_2$ is $\gamma_\textup{fock,vac} \approx 0.97$. We find that the probability of detecting vacuum in the second mode is $p_{0, \textup{fock}} \approx 52 \%$. For an input odd Schrödinger cat state, the purity after detecting vacuum in mode $v_2$ is $\gamma_\textup{cat,vac} \approx 0.93$, with probability $p_{0, \textup{cat}} \approx 43 \%$. For both cases, we see that the detection of vacuum purifies the quantum state in the mode and enhances the Wigner negativity as can be seen in Fig.~\ref{fig:heralded_quantum_state}.

\section{Conclusion} \label{sec:conclusion}
We have presented a formalism for calculating the transformation of an arbitrary pulsed input quantum state by a quadratic Hamiltonian, such as a parametric amplifier. Our method combines analytic and numerically efficient calculations, making it computationally tractable. The method involves three steps: i) decomposing the input-output operator relation into an effective set of modes, ii) finding the (squeezing) transformation of the input vacuum state within these effective modes, and iii) apply the transformed output mode operators to the squeezed vacuum state to obtain the output state in the desired modes. We provide a numerical implementation of the method and give examples of three different input states transformed by an optical parametric oscillator (OPO): a coherent state, a Schrödinger cat state and a Fock state. We observe how the output quantum state can be distributed over two entangled modes, and we have shown how detection of vacuum in one of the output modes may herald purification of the quantum state in the other output mode.

The method may be used for a wide range of systems besides OPO's, for instance optical parametric amplifiers or traveling wave parametric amplifiers. In this work we consider the input modes to be temporal/spectral modes, but the formalism may also be readily applied to, e.g., transverse spatial modes in 2D and 3D system platforms~\cite{chekhova-bright-squeezed-vacuum, chekhova-nanostructure-resonator, treps-spatio-temporal-squeezing}, Laguerre-Gaussian modes carrying orbital angular momentum \cite{gao-OAM, jing-OAM}, and relativistic spatio-temporal modes~\cite{kizmann-relativistic-squeezing}, to name a few. 

While our calculations rely heavily on the quadratic form of the Hamiltonian and the resulting effective treatment of Gaussian states through their first and second moments, we are able to deal with arbitrary input states. We imagine that it will also be possible to incorporate the effect of a single or a few non-Gaussian operations, such as further heralding upon detection among input or output modes, and possibly also address cumulant, second order mean field, approximations to the dynamics of systems with higher order nonlinearities.

\section*{Acknowledgment}
The authors acknowledge valuable discussions with Offek Tziperman.
V. Rueskov Christiansen acknowledges support from the Danish National Research Foundation through the Center of Excellence for Complex Quantum Systems (Grant agreement No. DNRF 152).
K. Mølmer acknowledges support from the Carlsberg Foundation through the ``Semper Ardens'' Research Project QCooL.
E. Hubenschmid acknowledges funding by the Deutsche Forschungsgemeinschaft (DFG) - Project No. 425217212 - SFB 1432.

\appendix
\renewcommand{\thesubsection}{\Alph{subsection}}
\renewcommand{\thesubsubsection}{\arabic{subsection}}

\section*{Appendix}

\subsection{The general output modes} \label{sec:calculating-v1-v2}
In this section we calculate an expression for the general two output modes $v_1$ and $v_2$. We start by noticing that the part of the $g_1(\omega, \omega')$-function in Eq. \eqref{eq:g1} that depends on the input state is fully described by the $f_u$ and $g_u$ functions. We can thus describe this part as a $2\times2$ matrix in this basis \cite{parametric-amplification-quantum-pulse}
\begin{align}
\begin{split}
   g_1^u &= |\alpha|^2 \zeta_u^2 f_u f_u^\dagger + \beta^* \zeta_u \xi_u g_u f_u^\dagger \\
   &+ \beta  \zeta_u \xi_u f_u g_u^\dagger + |\alpha|^2 \xi_u^2 g_u g_u^\dagger,
\end{split}
\end{align}
where $|\alpha|^2 = \braket{\a_u^\dagger \a_u}$ and $\beta = \braket{\a_u \a_u}$, and $f_u$ and $g_u$ denote vector representations of the mode functions. Since $f_u$ and $g_u$ functions are not orthogonal, we expand $g_u$ into parallel and orthogonal components to $f_u$, $g_u = \varepsilon f_u + \delta h_u$ where $\varepsilon = \braket{f_u, g_u}$ and $\delta = \sqrt{1 - |\varepsilon|^2}$. With this definition, we obtain the $2\times2$ matrix for $g_1^u$ in the $f_u$ and $h_u$ basis \cite{parametric-amplification-quantum-pulse}
\begin{align}\label{eq:matrix}
\begin{split}
    g_1^u = \left(\begin{matrix}
        |\alpha|^2( \zeta_u^2 + \xi_u^2 |\varepsilon|^2) + 2\zeta_u \xi_u \text{Re}(\beta^* \varepsilon) \\
        (\beta^* \zeta_u \xi_u + |\alpha|^2 \xi_u^2 \varepsilon^*) \delta
        \end{matrix} \right.
        \\
        \left.\begin{matrix}
        (\beta \zeta_u \xi_u + |\alpha|^2 \xi_u^2 \varepsilon) \delta \\ |\alpha|^2 \xi_u^2 \delta^2
    \end{matrix} \right).
\end{split}
\end{align}
The eigenvalue-decomposition of this matrix determines $v_1$ and $v_2$. The unnormalized eigenvectors for a $2\times2$-matrix $\begin{pmatrix}
    x & y \\ y^* & z
\end{pmatrix}$ are
\begin{align}
    v_\pm = \begin{pmatrix}
        x - z \pm \sqrt{(x - z)^2 + 4|y|^2}, & 2y^*
    \end{pmatrix}.
\end{align}
Inserting the values for the matrix elements in  \eqref{eq:matrix} we find
\begin{align}
    x - z &= |\alpha|^2 (\zeta_u^2 + \xi_u^2) + 2\zeta_u \xi_u \text{Re}(\beta^* \varepsilon) - 2|\alpha|^2\xi_u^2 \delta^2
\end{align}
and
\begin{align}
\begin{split}
    (x - z)^2 + 4|y|^2 &= (|\alpha|^2(\zeta_u^2 + \xi_u^2) + 2\zeta_u\xi_u \text{Re}(\beta^*\varepsilon))^2 \\ &+ 4 \zeta_u^2 \xi_u^2 \delta^2 (|\beta|^2 - |\alpha|^4),
\end{split}
\end{align}
which can be used to find explicit expressions for the normalized eigenvectors $v_1 = v_+ / ||v_+||$ and $v_2 = v_- / ||v_-||$, where the norm is
\begin{align}
\begin{split}
    ||v_\pm||^2 &= 2\Big[(x-z)^2 + 4|y|^2 \\
    &\pm (x-z)\sqrt{(x-z)^2 + 4|y|^2}\Big].
\end{split}
\end{align}

\subsection{Alternative proof for the restriction to two output modes} \label{app:alternative-two-mode-proof}
We offer here an alternative explanation for why a single mode input to the parametric amplifier leads to output in at most two modes in the output field. 
A general single mode input state $\ket{\psi}_u$ can be expanded in coherent states,
\begin{align}
    \ket{\psi}_u = \frac{1}{\pi} \int d^2\alpha \ket{\alpha}_u \braket{\alpha | \psi}_u = \frac{1}{\pi} \int d^2\alpha c_\alpha \ket{\alpha}_u
\end{align}
Expanding the input modes $u(\omega)$ in the eigenmode basis of the amplifier~\cite{Wasilewski2006},
\begin{equation}
    u(\omega) = \sum_j \eta_j \phi_j(\omega),
\end{equation}
which corresponds to writing each coherent state as a product of coherent states, 
\begin{align}
    \ket{\psi}_u = \frac{1}{\pi} \int d^2\alpha c_\alpha D(\alpha) \ket{0} = \frac{1}{\pi} \int d^2\alpha c_\alpha \prod_j D_j(\alpha_j) \ket{0},
\end{align}
where $D_j(\alpha_j)$ denotes a displacement operator acting on the eigenmode $\phi_j$, and $\alpha_j = \eta_j \alpha$. We insert the identity $S_j^\dagger (z_j) S_j(z_j)$ and transform $D_j(\alpha_j)$ with the squeezing operators
\begin{align}
\begin{split}
    S_j(z_j) D_j(\alpha_j) S_j^\dagger(z_j) = \exp&\left((\alpha_j\cosh{r_j} + \alpha_j^*\sinh{r_j}) \b_j^\dagger \right. \\
    &\left.-(\alpha_j^*\cosh{r_j} + \alpha_j\sinh{r_j}) \b_j\right).
\end{split}
\end{align}
Doing the product over $j$ leads to a sum over $j$ in the exponential. Since the factor in front of $\b_j$ and $\b_j^\dagger$ are the same except for the conjugate, we can define a mode
\begin{align}
    \b_\phi^\dagger = \frac{1}{\kappa} \sum_j \left(\alpha_j\cosh{r_j} + \alpha_j^*\sinh{r_j}\right) \b_j^\dagger,
\end{align}
where $\kappa$ ensures the normalization of the mode. We note that the mode $\phi(\omega)$ and its norm $\kappa$ depends on $\alpha$ and its conjugate. This displacement operator is equivalent to Eq.~\eqref{eq:displacement-operator}, and we note that it can be shown that $\phi = v$ and $\kappa = k$ from the main text. We continue with the notation used here and write
\begin{align}
    S\ket{\psi}_u = \frac{1}{\pi}\int d^2\alpha c_\alpha D_{\phi(\alpha)}[\kappa(\alpha)] \prod_j S_j(z_j) \ket{0}.
\end{align}
We consider a test mode $\b_s$. The mean number of photons in this mode is
\begin{align}
    \braket{\b_s^\dagger \b_s} = \left|\frac{1}{\pi}\int d^2\alpha c_\alpha \b_s D_{\phi(\alpha)}[\kappa(\alpha)] \prod_j S_j(z_j) \ket{0}\right|^2.
\end{align}
If $\b_s$ and $D_{\phi(\alpha)}$ commute, then $\b_s$ will only depend on squeezed vacuum contributions, as the displacement operator in mode $\phi(\alpha)$ will cancel with its adjoint. To commute, the mode $\b_s$ and $\b_\phi^\dagger$ must commute, which requires
\begin{align}
    \left[\b_s, \b_\phi^\dagger\right] = \frac{1}{\kappa} \sum_{jk} \sigma_k \left(\alpha_j\cosh{r_j} + \alpha_j^*\sinh{r_j}\right) \left[\b_k, \b_j^\dagger\right] = 0,
\end{align}
where we have expanded $\b_s = \sum_k \sigma_k \b_k$ in the eigenmode basis. The eigenmodes are orthonormal so this yields
\begin{align}
    \left[\b_s, \b_\phi^\dagger\right] = \frac{1}{\kappa} \sum_{j} \sigma_j \left(\alpha \eta_j\cosh{r_j} +  \alpha^*\eta_j^*\sinh{r_j}\right) = 0.
\end{align}
Since the $\phi$ mode depends on $\alpha$ and its conjugate, the $s$-mode must be orthogonal to the mode in the $2$-dimensional subspace spanned by $\alpha$ and $\alpha^*$ to be orthogonal to $\phi$ for any $\alpha$. We therefore see that in general, the output mode will span two modes.

\subsection{Mode decomposition of the output operator} \label{sec:output-decomposition}
In this appendix, we recall the mode decomposition in \cite{parametric-amplification-quantum-pulse}. We treat separately the general case leading to two relevant output modes and the special case leading to only a single relevant output mode.

\subsubsection*{Two output modes}
For the output mode $v_1(\omega)$, the output operator can be defined as
\begin{align}
    \b_{v_1} = \int d\omega v_1^*(\omega) \b(\omega).
\end{align}
Inserting the input-output relation in Eq. \eqref{eq:output-input-relation} yields
\begin{align}
    \b_{v_1} = \zeta^{v_1} \a_{f^{v_1}} + \xi^{v_1} \a_{g^{v_1}}^\dagger,
\end{align}
where we have defined modes $f^{v_1}$ and $g^{v_1}$ as
\begin{align}
    f^{v_1}(\omega) &= \frac{1}{\zeta^{v_1}} \int d\omega' F^*(\omega', \omega) v_1(\omega'), \\
    g^{v_1}(\omega) &= \frac{1}{\xi^{v_1}} \int d\omega' G^*(\omega', \omega) v_1^*(\omega'),
\end{align}
with normalization constants $\zeta^{v_1}$ and $\xi^{v_1}$. Note that these modes are not of the same form as $f_u$ and $g_u$ in Eqs.~(\ref{eq:f_u}, \ref{eq:g_u}), and therefore we use a different notation with raised indices, but they obey the relations $\zeta^v \braket{f^v, u} = \zeta_u \braket{v, f_u}$ and $\xi^v \braket{g^v, u} = \xi_u \braket{g_u, v}$ for any $u, v$. For the second mode $v_2$ we find a similar input-output relation
\begin{align}
    \b_{v_2} = \zeta^{v_2} \a_{f^{v_2}} + \xi^{v_2} \a_{g^{v_2}}^\dagger.
\end{align}
As we saw in App.~\ref{sec:calculating-v1-v2}, we can write the output modes as
\begin{align}
    v_1(\omega) &= C_1 f_u(\omega) + D_1 g_u(\omega), \\
    v_2(\omega) &= C_2 f_u(\omega) + D_2 g_u(\omega),
\end{align}
for some $C_i, D_i$, whose exact value is not relevant in the following. What is crucial is that under the insertion into $f^{v_i}$ and $g^{v_i}$ we get
\begin{align}
    f^{v_i}(\omega) &= \frac{C_i}{\zeta^{v_i}} f^{f_u}(\omega) + \frac{D_i}{\zeta^{v_i}} f^{g_u}(\omega), \\
    g^{v_i}(\omega) &= \frac{C_i^*}{\xi^{v_i}} g^{f_u}(\omega) + \frac{D_i^*}{\xi^{v_i}} g^{g_u}(\omega).
\end{align}
Using the fact that $F$ and $G^*$ must make the $\a_\textup{in}(\omega)$-mode obey the mode relation $[\a_\textup{in}(\omega), \a_\textup{in}(\omega')] = 0$, we find that \cite{Braunstein2005}
\begin{align}
    \int d\omega' F^*(\omega', \omega) G^*(\omega', \omega'') = \int d\omega' G^*(\omega', \omega) F^*(\omega', \omega'').
\end{align}
With this, we find that $\xi_u f^{g_u}(\omega) = \zeta_u g^{f_u}(\omega)$, and thus that the four $f^{v_i}$ and $g^{v_i}$ functions can be expressed by three orthonormal basis functions, which we label as $e_1, e_2$ and $e_3$. Furthermore, using $[\a_\textup{in}(\omega), \a_\textup{in}^\dagger(\omega')] = \delta(\omega - \omega')$, we show that \cite{Braunstein2005}
\begin{align}
\begin{split}
    &\int d\omega' \left(F^*(\omega', \omega) F(\omega', \omega'') - G^*(\omega', \omega) G(\omega', \omega'')\right) \\
    &= \delta(\omega - \omega''),
\end{split}
\end{align}
with which it can be found that $u(\omega) = \zeta_u f^{f_u}(\omega) - \xi_u g^{g_u}(\omega)$.
This means that the four $f^{v_i}$ and $g^{v_i}$ functions can be Gram-Schmidt decomposed into $u(\omega)$ and two other modes,
\begin{align}
\begin{split}
    &e_1 = u(\omega), \\
    &e_2 = t(\omega) = \frac{f^{v_1}(\omega) - \braket{u, f^{v_1}} u(\omega)}{\sqrt{1 - |\braket{u, f^{v_1}}|^2}}, \\
    &e_3 = s(\omega) = \frac{f^{v_2}(\omega) - \braket{u, f^{v_2}} u(\omega) - \braket{t, f^{v_2}} t(\omega)}{\sqrt{1 - |\braket{u, f^{v_2}}|^2 - |\braket{t, f^{v_2}}|^2}},
\end{split}
\end{align}
where we use $\braket{f, g} = \int d\omega f^*(\omega)g(\omega)$. Having obtained a basis for the modes, we can perform the decomposition of the output operators to yield
\begin{align}
\begin{split}
    \b_{v_i} &= \zeta^{v_i} \left(\braket{f^{v_i}, u} \a_u + \braket{f^{v_i}, t} \a_{t} + \braket{f^{v_i}, s} \a_{s}\right) \\
    &+ \xi^{v_i} \left(\braket{u, g^{v_i}} \a_u^\dagger + \braket{t, g^{v_i}} \a_{t}^\dagger + \braket{s, g^{v_i}} \a_{s}^\dagger\right),
\end{split}
\end{align}
for $i = 1, 2$. We have now obtained a decomposition into three modes. This allows us to write the decomposition in the matrix form used in Sec. \ref{sec:gaussian-variables-output-state},
\begin{align}
    \mathbf{\b} = \begin{pmatrix}
        E & F \\ F^* & E^*
    \end{pmatrix} \mathbf{\a}
\end{align}
with
\begin{align}
    E &= \begin{pmatrix}
        \zeta^{v_1} \braket{f^{v_1}, u} & \zeta^{v_1} \braket{f^{v_1}, t} & \zeta^{v_1} \braket{f^{v_1}, s} \\
        \zeta^{v_2} \braket{f^{v_2}, u} & \zeta^{v_2} \braket{f^{v_2}, t} & \zeta^{v_2} \braket{f^{v_2}, s}
    \end{pmatrix}, \\
    F &= \begin{pmatrix}
        \xi^{v_1} \braket{u, g^{v_1}} & \xi^{v_1} \braket{t, g^{v_1}} & \xi^{v_1} \braket{s, g^{v_1}} \\
        \xi^{v_2} \braket{u, g^{v_2}} & \xi^{v_2} \braket{t, g^{v_2}} & \xi^{v_2} \braket{s, g^{v_2}}
    \end{pmatrix}.
\end{align}

\subsubsection*{Single output mode}
We now consider the case where the input state feeds only into a single mode in the output field. This happens when the condition in Eq. \eqref{eq:single-mode-condition} is fulfilled for the input quantum state. The relevant output mode is identified in the main text as
\begin{equation}
    v(\omega) = \frac{\alpha \zeta_u f_u(\omega) + \alpha^* \xi_u g_u(\omega)}{k},
\end{equation}
A similar analysis as in the previous section shows that the $v$-mode is now spanned by two input modes, $e_1$ and $e_2$, and that the input mode in the single mode case can be written as
\begin{align} \label{eq:u-as-fv-gv}
    u(\omega) = \frac{k\zeta^v f^v(\omega) - k \xi^v g^v(\omega)}{\alpha},
\end{align}
so that the $f^v$ and $g^v$ functions can, in turn, be expanded on $u(\omega)$ and one other mode, which we can find by the Gram-Schmidt process
\begin{align}
\begin{split}
    &e_1 = u(\omega), \\
    &e_2 = t(\omega) = \frac{f^v(\omega) - \braket{u, f^v} u(\omega)}{\sqrt{1 - |\braket{u, f^v}|^2}}.
\end{split}
\end{align}
Finally, the output $\b_v$ operator can be written as
\begin{align}
\begin{split}
    \b_v &= \zeta^v \left(\braket{f^v, u} \a_u + \braket{f^v, t} \a_{t}\right) \\
    &+ \xi^v \left(\braket{u, g^v} \a_u^\dagger + \braket{t, g^v}\a_{t}^\dagger \right),
\end{split}
\end{align}
where we find using eq. \eqref{eq:u-as-fv-gv} that $\xi^v\braket{t, g^v} = \zeta^v \braket{t, f^v}$ and furthermore that $\braket{t, f^v} = \sqrt{1 - |\braket{u, f^v}|^2}$, so the $E$ and $F$ matrices are
\begin{align}
    E &= \begin{pmatrix}
        \zeta^v \braket{f^v, u} & \zeta^v \sqrt{1 - |\braket{u, f^v}|^2}
    \end{pmatrix}, \\
    F &= \begin{pmatrix}
        \xi^v \braket{u, g^v} & \zeta^v \sqrt{1 - |\braket{u, f^v}|^2}
    \end{pmatrix}.
\end{align}

\subsection{Mean values and covariance matrix for the squeezed vacuum and squeezed coherent state} \label{sec:vacuum-state-covariance}
In this appendix, we derive the covariance matrix of the squeezed vacuum state in a relevant set of modes. We begin with the expression of the input-output relation in terms of matrices $E$ and $F$ in Eq. \eqref{eq:input-output-matrix}. We then go to the quadrature basis
\begin{align}\label{eq:Amult}
    \begin{pmatrix}
        \mathbf{\q}_\textup{out} \\ \mathbf{\p}_\textup{out}
    \end{pmatrix} = A \begin{pmatrix}
        \mathbf{\q}_\textup{in} \\ \mathbf{\p}_\textup{in}
    \end{pmatrix}
\end{align}
where $\mathbf{\q_\textup{in}}$ and $\mathbf{\p_\textup{in}}$ are vectors with the $\q_i = (\a_i + \a_i^\dagger)/\sqrt{2}$ and $\p_i = (\a_i - \a_i^\dagger)/i\sqrt{2}$ quadratures as entries and similar for $\mathbf{\q_\textup{out}}$ and $\mathbf{\p_\textup{out}}$ in terms of $\b_i$, and where $A$ is the $2N\times2M$ matrix
\begin{align}
    A = \begin{pmatrix}
        \Re(E + F) & \Im(F - E) \\
        \Im(E + F) & \Re(E - F)
    \end{pmatrix}.
\end{align}
Equipped with this transformation, we can find the second-order moments for the output state
\begin{align}
    \braket{\begin{pmatrix}
        \mathbf{\q}_\textup{out} \\ \mathbf{\p}_\textup{out}
    \end{pmatrix} \begin{pmatrix}
        \mathbf{\q}_\textup{out} \\ \mathbf{\p}_\textup{out}
    \end{pmatrix}^T} = A \braket{\begin{pmatrix}
        \mathbf{\q}_\textup{in} \\ \mathbf{\p}_\textup{in}
    \end{pmatrix} \begin{pmatrix}
        \mathbf{\q}_\textup{in} \\ \mathbf{\p}_\textup{in}
    \end{pmatrix}^T} A^T
\end{align}
The input vacuum state has the matrix of second order moments,
\begin{align}
    \braket{\begin{pmatrix}
        \mathbf{\q}_\textup{in} \\ \mathbf{\p}_\textup{in}
    \end{pmatrix} \begin{pmatrix}
        \mathbf{\q}_\textup{in} \\ \mathbf{\p}_\textup{in}
    \end{pmatrix}^T} = \frac{1}{2} \begin{pmatrix}
        \mathbf{I}_N & i \mathbf{I}_N \\ -i \mathbf{I}_N & \mathbf{I}_N
    \end{pmatrix},
\end{align}
and the output quadrature covariance matrix with elements, $\sigma_{kl} = \braket{\{\hat{r}_k, \hat{r}_l\}} - 2\braket{\hat{r}_k}\braket{ \hat{r}_l}$ becomes
\begin{align}
    \bm{\sigma}_\textup{out} = \mathbf{I}_4 + 2 \begin{pmatrix}
        \text{Re}(EF^T + FF^\dagger) & \text{Im}(EF^T - FF^\dagger) \\
        \text{Im}(EF^T + FF^\dagger) & \text{Re}(FF^\dagger - EF^T)
    \end{pmatrix},
\end{align}
where we have used the relations $EE^\dagger - FF^\dagger = I$, $EF^T = FE^T$ \cite{bloch-messiah}.

The case of a squeezed coherent state, occupying a single mode in the output, leads to the same covariance matrix as above, since the covariance matrix of a coherent state is equal to that of vacuum. For the single mode transformation, this is
\begin{align}
   \bm{\sigma}_\textup{vac,out} = \begin{pmatrix}
       4 (\zeta^v)^2 + \frac{|\alpha|^2}{k^2} - 4 \frac{\zeta_u}{k} \textup{Re}(z) & -2\frac{\zeta_u}{k} \textup{Im}(z) \\
       -2\frac{\zeta_u}{k} \textup{Im}(z) & \frac{|\alpha|^2}{k^2}
   \end{pmatrix},
\end{align}
where we have defined $z = \alpha \braket{v, f_u}$, and where $\zeta^v$ can be found in appendix \ref{sec:output-decomposition}.
The input mean displacement vector for a coherent state is $(\sqrt{2}\Re(\alpha),0,\sqrt{2}\Im(\alpha),0)^T$, and multiplying by $A$ for the single-mode transformation yields the displacement vector for the output mode, cf., Eq.~\eqref{eq:Amult}
\begin{align}
   \mathbf{\bar{r}} = \sqrt{2}\begin{pmatrix}
       2\zeta_u \textup{Re}(z) - \frac{|\alpha|^2}{k} \\
       0
   \end{pmatrix}.
\end{align}

\bibliography{bibliography.bib}
\end{document}